# Synthesis of nanocrystalline Spinel phase by Mechanical Milling of Al-Cu-Fe and Al-Cu-Cr-Fe Quasicrystalline alloys

T. P. Yadav[1] N. K. Mukhopadhyay[1], R. S. Tiwari[1], and O. N. Srivastava[1]

[1]Department of Physics, Banaras Hindu University, Varanasi 221005, India
[2]Department of Metallurgical Engineering, Institute of Technology,
Banaras Hindu University, Varanasi 221005, India

## Abstract

In the present study, low-temperature synthesis of $(Cu,Fe)Al_2O_4$ and $(Cu,Cr,Fe)Al_2O_4$ spinels from the quasicrystalline phases was investigated with the variation of process parameters during milling and annealing. The milling of the quasicrystalline materials was carried out in an attritor mill at 400 rpm for 40 hours with ball to powder ratio of 40:1 in hexane medium. Subsequently, annealing was performed in an air ambience for 10, 20, and 40 h at 400, 500, 600 and 700 °C in side a furnace in order to oxidize the mechanically milled quasicrystalline phase for the possible formation of the spinel phase. It was found that after annealing at higher temperature (~500 °C), mechanically milled quasicrystalline alloy transformed to spinel phase whereas annealing at lower temperature (<500 °C) it led to the formation of B2 phase as a major one along with minor amount of oxide phase. The X-ray diffraction and transmission electron microscopy of the annealed samples confirmed the formation of spinel phase with an average grain size of ~20-40 nm. It is interesting to note that the nanospinel phases showed the different colors during various annealing time and temperature. The optical properties of nanospinel materials, investigated employing UV visible spectrometer exhibited absorption characteristics.

**E-mail: hepons@yahoo.com  (ONS)**
**yadavtp@gmail.com  ( TPY)**

# 1. Introduction

Quasicrystals (QCs) are orientationally ordered structures which may often possess classically forbidden rotational symmetries (e.g. five-fold, eight-fold, ten-fold and twelve- fold). Due to the aperiodicity, these materials are expected to exhibit properties that are very different from conventional metallic materials [1-2] and these properties can be exploited for industrial applications [3-6]. More recently, there have been studies on the formation of spinel phase from quasicrystalline materials in Al-Co-Ni and Al-Ni-Fe systems [7-8]. The synthesis of aluminum–transition metal nanocrystalline spinel has been investigated intensively in the recent years due to their unique potential applications. The spinel structure is characterized by a simultaneous occupation of tetrahedral and octahedral positions by metallic cations in an FCC oxygen lattice. Aluminum based spinels constitute an important class of advanced ceramic materials with a wide variety of technological applications [9]. The $CoAl_2O_4$, $NiAl_2O_4$ and, $CuAl_2O_4$, possess interesting electronic, magnetic and catalytic properties and are used in industry as ceramic pigments, coatings and catalysts [10-11]. The spinel materials can be prepared by many methods such as solid-state reaction, hydrothermal, alkoxide hydrolysis, sol-gel method and microwave-induced methods [12-14]. The most general method is the solid-state reaction, which involves the mixture of metal oxides followed by sintering at high temperature. Recently, the nanospinel materials have been synthesized milling quasicrystalline precursor, followed by annealing [7, 8, & 15]. It is not clear in any of the prior work whether true thermodynamic equilibrium of the quasicrystal and spinel phases could be achieved. Here we reported the formation of spinel phase from Al-Cu-Fe icosahedral and Al–Cu–Cr-Fe decagonal quasicrystals by mechanical milling of as cast alloys and annealing in air ambient. We showed on the basis of present investigation that the Al–Cu–Cr-Fe decagonal quasicrystal is more suitable precursor for the formation of single spinel phase. The optical characteristics were also studied.

# 2. Experimental Procedure

The alloy with a composition close to $Al_{65}Cu_{20}Fe_{15}$ and $Al_{65}Cu_{20}Cr_8Fe_7$ (Al=99.98%, Cu=99.99%, Cr=99.97%, Fe = 99.98 % pure) was prepared by melting in an R.F. induction furnace under a dry argon atmosphere. The individual elements were at first mixed in correct stoichiometric proportions and pressed into a cylindrical pellet of 2 cm diameter and 0.75 cm thickness by applying a pressure of ~ $3 \times 10^4$ N/m$^2$. The pellet (5 g by weight) was then placed in a silica tube surrounded by an outer Pyrex glass jacket. Under continuous flow of argon gas into the silica tube, the pellet was melted using radio frequency induction furnace (18 kW). The ingots formed were re-melted several times to ensure homogeneity. This pre-alloyed as cast ingot was crushed to particles less than 0.5mm in size and placed in an attritor ball mill (Szegvari Attritor) with ball to powder ratio of 40:1. The attritor has a cylindrical stainless steel tank of inner diameter 13 cm. The grinding balls made of stainless steel are of 6 mm in diameter. The speed of the mill was maintained at 400 rpm. The milling operation was conducted for various time ranging from 5h to 40h using hexane as a process control agent (PCA). Powders obtained after 40 h of milling was further annealed at 400, 500, 600 and 700°C for time ranging from 10 to 60h in air ambient. The air annealing has been done for oxidation of milled powders to form spinel phase. Structural and microstructural characterizations of the mechanically activated and annealed powders were performed using X-ray diffraction (XRD) with CuK$_\alpha$ radiation (Philips PW-1710 X-ray diffractometer, λ =1.54026 Å), scanning electron microscopy (Philips XL-20) and transmission electron microscopy (TEM- Techni G$^{20}$ at 200kV).The effective crystallite size and relative strain of mechanically milled powders as well as heat-treated products were calculated based on line broadening of XRD peaks. The use of the Voigt function for the analysis of the integral breadths of broadened X-ray diffraction line profiles forms the basis of a single line method of crystallite-size and strain determination [16].

## 3. Results and discussions

Fig. 1(a) shows XRD pattern of the icosahedral phase obtained from the as cast Al$_{65}$Cu$_{20}$Fe$_{15}$ alloy. All of the peaks are from the icosahedral phase; no other phase could be identified. Fig. 1(b) shows the XRD pattern from 40h mechanical milled (MM) powder, exhibiting broadening of the peaks belonging to the B2 phase. From the XRD pattern it is evident that the initially sharp diffraction peaks are considerably broadened after ball milling due to a decrease in the particle size and an increase in internal lattice strains. The effect of annealing in air of MM alloy can be seen from fig. 1 (curve c) where it shows the spinel phase as a major constituent phase along with aluminum oxides (Al$_2$O$_3$) which could not be avoided even after annealing for more time.

In order to investigate the effect of the Cr substitution in place of Fe for the formation of spinel structure, the quaternary alloy showing the decagonal phase was selected. Fig. 2(a) shows

XRD pattern of the decagonal phase obtained from the as cast $Al_{70}Cu_{20}Cr_8Fe_7$ alloy. The diffraction pattern corresponding to decagonal phase can be indexed by using six independent indices as proposed by Mukhopadhyay and Lord [17]. All of the peaks are from the decagonal phase, no other phase was identified. Fig. 2(b) shows the XRD pattern from 40 h MM powder exhibiting broadening of the peaks belonging to the B2 phase. The investigation of the formation of spinel phase from MM decagonal alloy has been carried out by annealing in air. In this case no other phase has been observed. In comparison to the $Al_{65}Cu_{20}Fe_{15}$ sample, the formation of single phase spinel has been observed in $Al_{65}Cu_{20}Cr_8Fe_7$ sample. The annealing has been done in air, a significant phase transformation has been found to occur in case of 40 h mechanically activated quasicrystalline powder. XRD patterns (fig 1c & fig. 2c) show the peaks corresponding to the spinel structure with the lattice parameter 8.1 ± 0.04 Å and 8.05 ±0.04 Å respectively. The compositional analysis of 40 h MM and air-annealed powders of both the alloy was analyzed by EDX attached to a high-resolution transmission electron microscope. After annealing in air the amount of oxygen increases up to 64.7 at % in case of $Al_{65}Cu_{20}Fe_{15}$ and 50.8 at % in case of $Al_{65}Cu_{20}Cr_8Fe_7$, indicating the amount of oxygen pick up from air. It is important to point out that those quasicrystals particles (without milling) after annealing in air for extended period could not yield the formation of the spinel phase except some individual oxide phases. Thus the milling of the quasicrystals is absolutely important in order to form the spinel phase as it enhances the kinetics of the formation. It is obvious that the high degree of oxidation has caused the formation of spinel structures from the MM materials. It is understood that after 40h of milling, the surface and interface areas are increased due to the formation of smaller particles as well as the nanocrystallites. Thus the kinetics of oxidation rate, which is controlled by oxygen diffusion, becomes faster. Since oxygen affinity of Cr is higher than Fe, the spinel phase in Al-Cu-Cr-Fe alloy was found to be formed very easily.

The rigorous transmission electron microscopic (TEM) investigation by obtaining selected area diffraction patterns and microstructural features at different stages have been carried out. Fig. 3 shows TEM investigation of 40h MM and air annealed $Al_{65}Cu_{20}Fe_{15}$ and $Al_{64}Cu_{29}Cr_8Fe_7$ alloys. The bright field microstructure of 40 h MM and 60 h air annealed sample of $Al_{65}Cu_{20}Fe_{15}$ can be seen in Fig. 3(a). A very fine equiaxed particles of the nanometer size (20~30 nm) are visible. The corresponding electron diffraction pattern, shown in Fig. 3(b) was indexed due to spinel phase. Fig. 3 (c) shows a bright field TEM micrograph of $Al_{65}Cu_{20}Cr_8Fe_7$ 40 h MM alloy followed by annealing at 600 °C for 60 h in air. The large number of small size particles embedded in the big grain can easily be discerned in the figure. The average particle size is of 30~40 nm. The corresponding electron diffraction patterns are shown in Fig.3 (d). All the diffraction rings are indexed in the terms of $(Cu,Cr,Fe)Al_2O_4$ spinal phase with lattice parameter of 8.1Å of an FCC. This lattice parameter agrees reasonably well with the XRD results. From the present investigation, it is clear that the formation of spinel phase during air annealing is complete as no existence of other quasicrystalline/ crystalline phases could be identified.

To find out the optical properties of MM and air annealed samples, the UV-vis (ultraviolet- visible) spectroscopy experiments have been carried out. Fig. 4 shows the UV–vis spectra of the as-grown spinel phase from $Al_{65}Cu_{20}Fe_{15}$ and $Al_{65}Cu_{20}Cr_8Fe_7$ materials. It is clear from the spectra that these spinel phases are really sensitive to the UV-vis radiation. It can be pointed out that the quasicrystalline or nano B2 phase did not show any color changes or any absorption in their corresponding optical spectra, suggesting that they were optically insensitive phases. $Al_{65}Cu_{20}Fe_{15}$ spinel material showing light brown colour, exhibited a small but sharp absorbance peak at 760nm, whereas $Al_{65}Cu_{20}Cr_8Fe_7$ spinel material showing the dark brown color exhibited a broader peak at 760nm. It can be further stated that the overall amount absorbance in the whole UV–vis range, is somewhat more in case of $Al_{65}Cu_{20}Cr_8Fe_7$ spinel phase, which could be responsible for dark brown color. The sharp peak in $Al_{65}Cu_{20}Fe_{15}$ spinel may suggest that the tetrahedrally coordinated ions are much more ordered in nature compared to the other one reflecting the existence of discrete band structures due to the electronic transition allowed by spin. As a result somewhat broader peak is observed in $Al_{65}Cu_{20}Cr_8Fe_7$ spinel phase due to its diffuse band structure arising from the disordering among the Cr/Fe ions [18]. However it is worth pursuing to investigate this aspect in order to understand the detailed electronic transitions responsible for such optical characteristics. It may be pertinent to mention that this kind of behavior has been interpreted as a characteristic for tetrahedral transitions allowed by the spin and causing for the blue color in the case of $CoAl_2O_4$ spinel [19].

## 4. Conclusions

The $(Cu,Fe)Al_2O_4$ and $(Cu,Cr,Fe)Al_2O_4$ spinels were synthesized from icosahedral and decagonal quasicrystalline precursors using mechanical milling and followed by air annealing. The minimum annealing temperature and time were found to be 600 °C and 60h respectively. In both the cases the evolution of the spinel structure started upon air annealing at 600 °C from the B2 phase formed due to milling of quasicrystalline phases for 40h. From the analysis it appeared that Al-Cu-Cr-Fe decagonal phase exhibits better spinel forming ability compared to that of Al-Cu-Fe icosahedral phase. The spinel phase formation by annealing of the as-cast quasicrystal phase (without milling) in air for extended period of time was not found to be feasible. Thus the milling was found to be absolutely necessary to obtain the spinel phase. The UV-vis spectra suggested that the spinel phase were sensitive to the optical radiation and showed slightly different optical characteristics which can be attributed to the Cr and Fe ions in the structure.

*Acknowledgement*

The authors would like to thank Prof G.V.S. Sastry, Prof. R.K. Mandal, Prof. B.P. Asthana, Prof A.S.K.Sinha and Dr. M.A.Shaz, for many stimulating discussions.

## Reference

[1] Archambault, P.; Janot,C .: Thermal conductivity of quasicrystals and associated processes. Mater. Res. Soc. Bulletin. **22**(11) (1997) 48-53.

[2] Mancinelli, Chris.;. Ko1, J. S.; Jenks, C.J .;. Thiel, Patricia A.; Ross, Amy R.. Lograsso, Thomas A, . Gellman. Andrew J. : Comparative Study of the Tribological and Oxidative Properties of AlPdMn Quasicrystals and Their Cubic Approximants. Mat. Res. Soc. Symp. Proc. **643** ( 2001) K8.2.1- K8.2.10

[3] Rónán, McGrath.; Julian Ledieu.; Erik, J. Cox,; Renee, D. Diehl.: Quasicrystal surfaces: structure and potential as templates. J. Phys.: Condens. Matter. **14** (2002) R119-R144

[4] Besser, M.F.; Eisenhammer, T.: Deposition and Applications of Quasicrystalline Coatings. Mater.Res. Soc. Bulletin, **22** (1997) 59-63

[5] Urban, K.; Feuerbacher, M..; Wollgarten, M..: Mechanical Behavior of Quasicrystals Mater.Res. Soc. Bulletin.**22** (1997) 65-68

[6] Kelton,K.F.; Gibbons, P.C.: Hydrogen Storage in Quasicrystals., Mater.Res. Soc. Bulletin. 22 (1997) 69-72

[7] Yadav, T.P.; Mukhopadhyay, N.K.; Tiwari, R.S.; Srivastava,O.N.: Synthesis of nanocrystalline (Co,Ni)Al$_2$O$_4$ spinel powder by mechanical milling of quasicrystalline material, Journal of Nano-Science and Technology. **7** (2007) 575-579

[8] Yadav, T.P.; Mukhopadhyay, N.K.; Tiwari, R.S.; Srivastava,O.N.: Low-temperature synthesis of nanocrystalline spinel by mechanical milling and annealing of Al-Ni-Fe decagonal quasicrystals., Phil. Mag. 2008 ( In press)

[9] Meyer, F.; Hempelmann, Rolf.; Mathur, S.; Veithb, Michael.: Microemulsion mediated sol–gel synthesis of nano-scaled MAl2O4 (M=Co, Ni, Cu) spinels from single-source heterobimetallic alkoxide precursors, J. Mater. Chem. **9** (1999) 1755-1763

[10] Woelk, H.J,; Hoffmann, B.; Mestl, G.; Schloegl, R .: Experimental archaeology: Investigation on the copper-aluminum-silicon-oxygen system., J. Am. Ceram. Soc. **85** (2002) 1876-1878.

[11] Shimizu, K.; Maeshima, H..;Yoshida, H, Satsuma, A.; Hattori,T.: Spectroscopic Characterization of Cu-Al2O3 Catalyst for Selective Catalytic Reduction of NO with Propene., Phys. Chem. Chem. Phys. **2** (2000) 2435-2439.

[12] Cho, W.S .; Kakihana, M .: Crystallization of ceramic pigment CoAl$_2$O$_4$ nanocrystals from Co–Al metal organic precursor. Journal of Alloys and Compounds **287** (1999) 87-90,

[13] Jeevanandam, P.; Koltypin Yu. Gedanken, A .: Preparation of nanosized nickel aluminate spinel by a sonochemical method. Mater Sci Eng. *B* **90** (2002)125-232 .

[14] Peelamedu, R.; Agrawal. D.; Roy, R.:Microwave induced reaction sintering of Ni aluminates. Mat. Lett*.* **55** (2002) 234-240

[15] Yadav, T.P.; Mukhopadhyay, N.K.; Tiwari, R.S.; Srivastava,O.N.: Evolution of a nanocrystalline (Co,Ni)Al$_2$O$_4$ spinel phase from quasicrystalline precursor, Journal of Applied Ceramic Technology (2008) (on line available DOI:10.1111/j.1744-7402.2008.02243.x)

[16] Yadav,T.P., Mukhopadhyay, N.K., Tiwari, R.S.; Srivastava,O.N .: Studies on the formation and stability of nano-crystalline Al$_{50}$Cu$_{28}$Fe$_{22}$ alloy synthesized through high-energy ball milling", Mater. Sc. Eng. A **393** (2005) 366-373

[17] Mukhopadhyay N.K .; Loard, E.A.: Least path criterion (LPC) for unique indexing in a two-dimensional decagonal quasicrystal. Acta. Cryst. A. **58** (2002) 424-428

[18] Gaffiney, E.S.: Spectra of Tetrahedral $Fe^{2+}$ in $MgAl_2O_4$. Phys. Rev. B **8** (1973) 3484-3486.

[19] Merikhi, J.; Jungk, H.O.; Feldmann, C. : Sub-micrometer $CoAl_2O_4$ pigment particles — synthesis and preparation of coatings.; J. Mater. Chem. **10** (2000) 1311-1314

**Figure Captions:**

Fig.1   X-ray diffraction pattern obtained from the  as cast  $Al_{65}Cu_{20}Fe_{15}$ alloy showing the formation of icosahedral phase (curve 'a')  and  40 h mechanically milled powder (curve 'b'), peak broadening corresponding to (110) ( $2\theta \sim 43°$)   curve indicates the formation of nano B2  phase. The air annealing at $600^oC$ for 60h of milled powders (curve c) shows the formation of spinel phase along with a minor amount of $Al_2O_3$.

Fig. 2   X-ray diffraction pattern obtained from as cast  $Al_{64}Cu_{20}Cr_8Fe_7$ alloy ,  showing the formation of decagonal phase  (curve 'a') and the powder after mechanical milling for 40 h (curve `b `) and subsequent annealing at 600° C for 60h in air (curve 'c')  . Curve 'c' indicates the formation of nanospinel phase. No other oxide phase can be detected.

Fig.3   (a) Bright field TEM image of powder sample after 40 h milling followed by air annealing at 600 °C of $Al_{65}Cu_{20}Fe_{15}$ alloy  showing the of nano particle   of the order  of 20-40 nm in size,  (c) the bright field TEM image of  40 h MM and air annealed at 600 °C  for 60h of  $Al_{65}Cu_{20}Cr_8Fe_7$ alloy.

Fig.4   UV–visible spectra of the powders obtained by 40 h milling followed by air   annealing at 600 °C of  (a)$Al_{65}Cu_{20}Fe_{15}$ alloy  and(b)$Al_{64}Cu_{20}Cr_8Fe_7$ alloy , showing a small peat at 760nm, which is sharper in (a) but broader in (b).

Fig.1

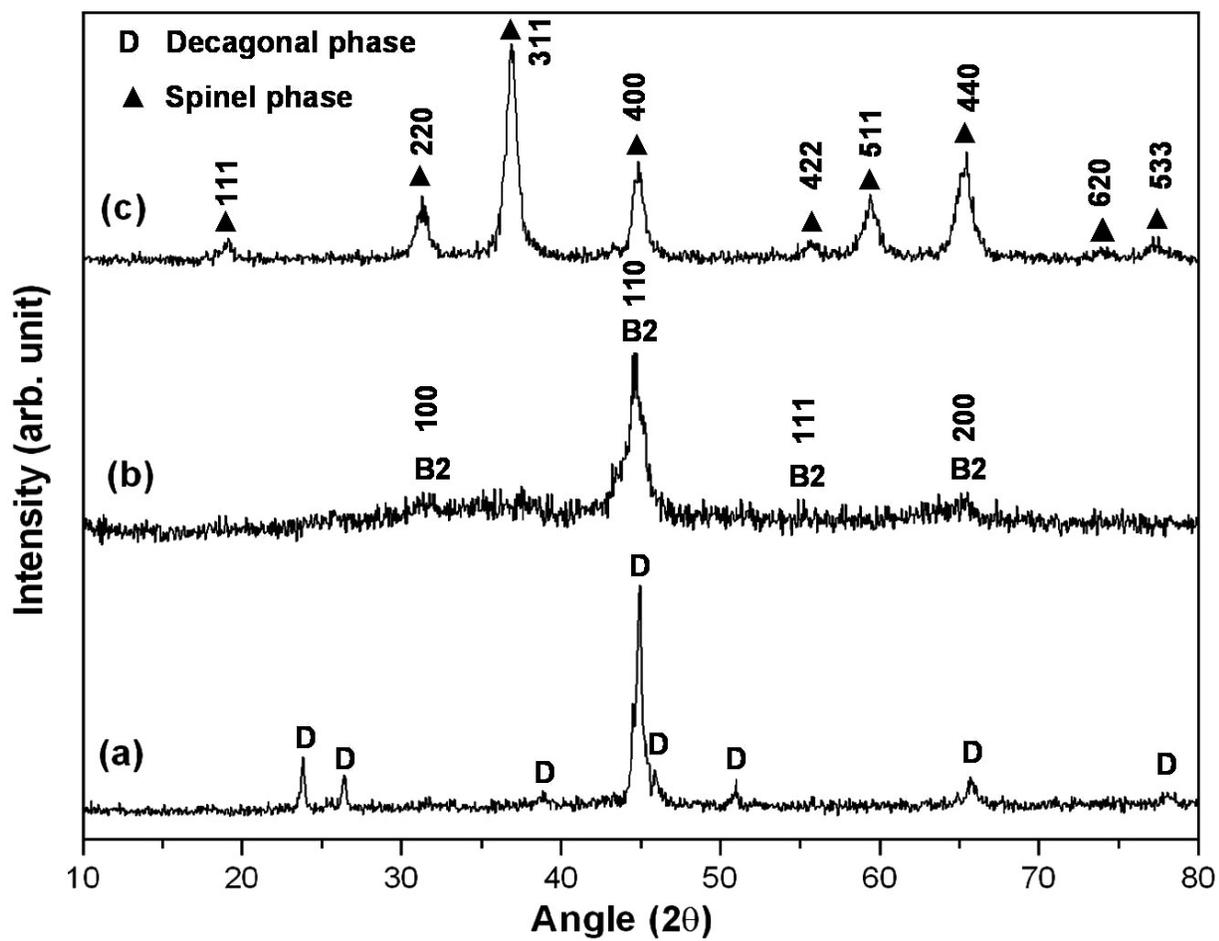

**Fig.2**

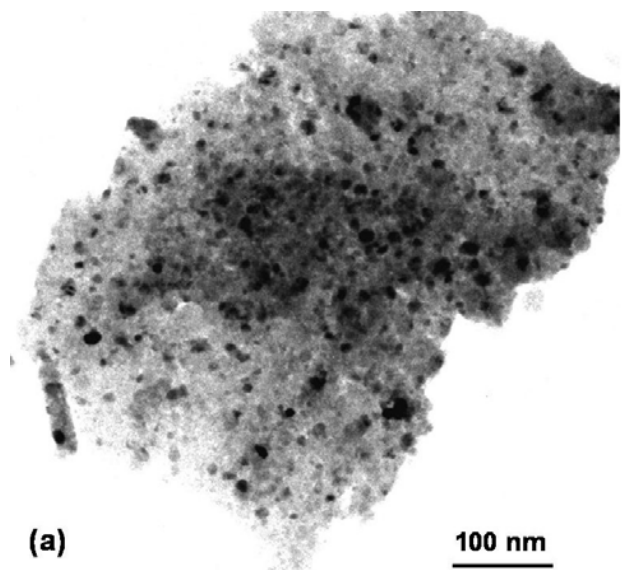 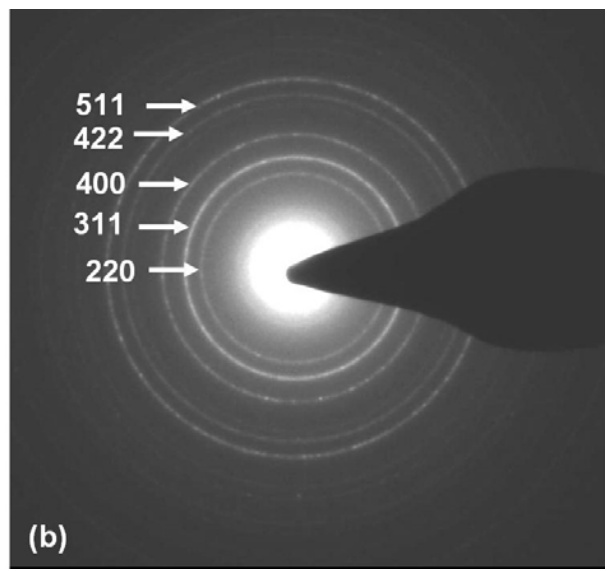

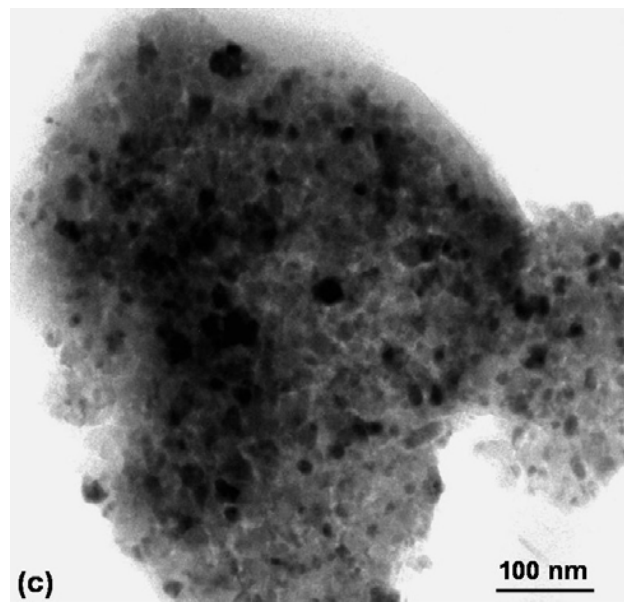 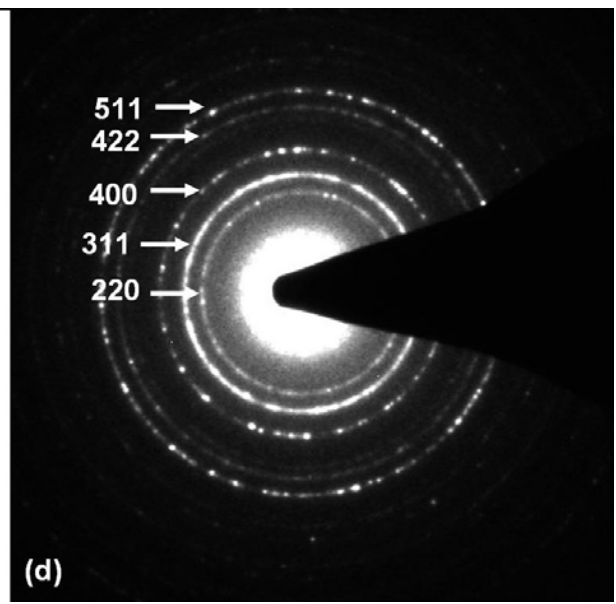

**Fig.3**

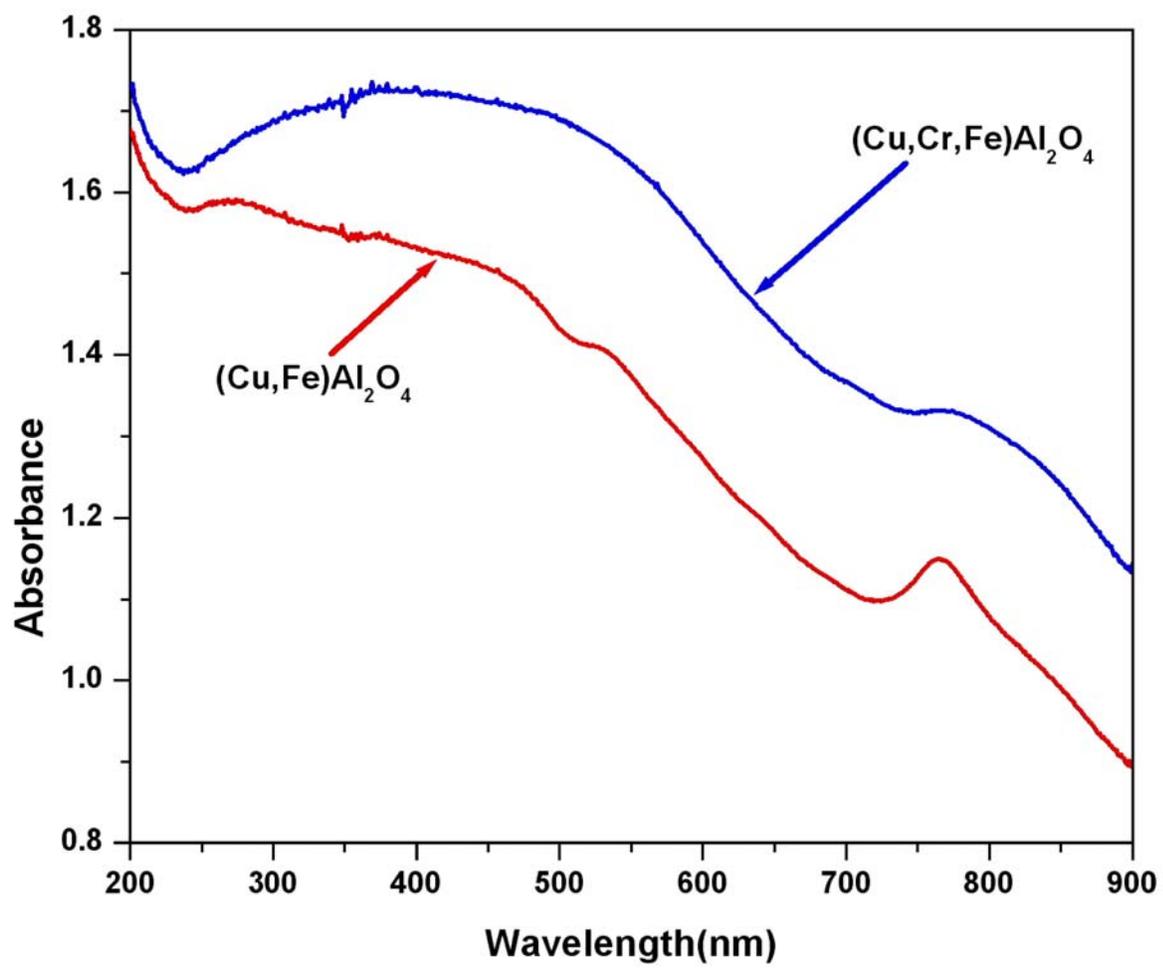

**Fig.4**